\documentclass{article}
\usepackage{graphicx}
\usepackage[margin=1.0in]{geometry}

\usepackage{amsmath}
\usepackage{amsfonts}
\usepackage{ dsfont }
\usepackage{ dsfont }
\usepackage{ amssymb }
\usepackage{authblk}
\usepackage{braket}
\usepackage{float}
\usepackage[style=ieee, citestyle=numeric-comp, backend=biber]{biblatex}

\begin{document}

\title{A Repulsive Casimir Effect for Circular Geometries}
\author[1]{Daniel Davies\thanks{Email: dadavies@ucsc.edu}}
\date{\today}
\affil[1]{\small{Department of Physics, University of California, Santa Cruz, Santa Cruz CA 95064}}

\maketitle

\begin{abstract}
Using numerical analytic continuation, we compute the Zeta function for the Casimir Effect for circular geometries in 2+1 dimensions. After subtraction of the simple pole of the zeta function, essentially MS renormalization, we find the Casimir force is repulsive. Implications for the stability of 2+1 dimensional domain walls and Axion membranes in 3+1 dimensions are discussed, especially in the context of strongly coupled underlying physics.
\end{abstract}


\section{Background \& Results}

In quantum field theories, extended objects play a critical role in phase transitions and often appear relevant in strong coupling limits. Many forms of these objects are known, and much has been detailed of their classical properties: tension, stability, etc. However, one might wonder whether at small distance scales or at large coupling the classical details of these objects are modified significantly by quantum corrections. Especially so when these limits coincide. The first and most studied of these quantum corrections is the Casmir Effect, or in path integral language, the evaluation of the functional determinant of a fluctuation operator and its contribution to the effective action.\\

Of particular interest to this work is the case of a circular, or disk geometry. We will discuss some physically relevant examples at the end of this work, but for now the problem statement is the following: calculate the infinite sum

\begin{align}
    E_\text{Cas} = \sum_i \omega_i
\end{align}

\noindent where $\omega$ are the energies of 2-dimensional waves that have some Dirichlet or Neumann boundary conditions at finite radius $r=a$ (here we focus on the Dirichlet case). Assuming such waves are massless, the energies are given by the zeros of the Bessel functions of the first kind, and the sum to be taken over all such zeros (there are countably infinite of them).

\begin{align}
   E_\text{Cas}(a) = \sum_i \omega_i &= \frac{1}{a} \sum_{\ell=-\infty}^\infty \sum_{n=1}^\infty j_{\ell,n}\\
   \text{where}\quad J_\ell(j_{\ell,n}) &= 0, \quad n = 1,2,3,\cdots
\end{align}

This summation is obviously divergent, but the correct quantity to evaluate is the Zeta function of this series, analytically continued to a region of the complex plane where it differs from the summation. In other words:

\begin{align}
E_\text{Cas}(a) &=a^{-1} \zeta_\odot(-1)\\
\zeta_\odot(s) &=  \sum_{\ell=-\infty}^\infty \sum_{n=1}^\infty \left(j_{\ell,n}\right)^{-s}
\end{align}

\noindent for $\text{Re}(s) > 2$. Much is known about this Zeta function, including that it possesses simple poles at $s=2,1,-1,-3,\cdots$\cite{1}. That is to say, the Casimir Energy is, strictly speaking, infinite. However, this is not an infrared divergence and cannot be traced back to an infinite volume limit. It is a UV problem associated with the asymptotic spacing of the zeros $j_{\ell,n}$. It is known from McMahon's asymptotic expansion that for fixed $\ell$ and large $n$, we have\cite{2}\\

\begin{align}
    j_{\ell,n} \sim \pi\left(n+\frac{\ell}{2} + \frac{1}{4}\right) - \frac{4\ell^2-1}{8\pi\left(n+\frac{\ell}{2} + \frac{1}{4}\right)} - \mathcal{O}\left(\frac{1}{n^3}\right)
\end{align}

\noindent Already it is obvious that the summation over $n$ produces a pole from the Hurwitz Zeta function, i.e. the second term. However, The remaining summation over $\ell$ is not finite either, having its own logarithmic divergences. The result is a simple pole at $s=-1$. Thankfully, the residue of this pole (and the ones at $s=2,1,-3$) are known\cite{1}. The Zeta function of interest has the form:

\begin{align}
    \zeta_\odot(s) = \frac{1}{2(s-2)}-\frac{1}{2(s-1)}-\frac{1}{128(s+1)} + \varphi(s)
\end{align}

\noindent where $\varphi(s)$ is analytic for $\text{Re}(s) > -3$. The residue at $s=-3$ is quite small, being $111/2^{15}$. We can therefore define the MS renormalized Casimir Energy as follows:

\begin{align}
    E^{\text{ren}}_\text{Cas}(a) &= a^{-1}\times \lim_{s\rightarrow -1}\left(\zeta_\odot(s) + \frac{1}{128(s+1)}\right)\\
    &= \frac{1}{a}\left(\frac{1}{12}+\varphi(-1)\right)
\end{align}

This author is unaware of any previous numerical determination of $\varphi(-1)$, but we have found the value to be approximately $-0.07594$, with relative error in the 5\% range. This easily places the coefficient in equation (8) at the positive value $\approx 0.0074 \pm 0.0038$, indicating that the renormalizaed quantum forces for this geometry are repulsive, and act to increase the radius. This is in contrast to the Casimir forces in 1 dimension, which are attractive and have coefficient $-1/12$.

\section{Methods}

\subsection{Data Generation}

Efficient computation of the $j_{\ell,n}$ for small $\ell$ and $n$ is handled quite well by libraries like Python's SciPy, which utilizes Newton's algorithm for root finding. For large $n$ and small $\ell$, the McMahon asymptotic expansion is preferable to root finding. The built-in SciPy root finder for $J_\ell(x)$ uses $x=0$ as its starting point however, which poses a significant problem for larger values of $\ell$. These functions are extremely small near the origin for larger values of $\ell$, and do not grow to appreciable values until $x$ is comparable with the first root $j_{\ell,1}$. For example, $J_{200}(x)$ is smaller than about 1 part in $10^{15}$ until $x \sim 150$. Floating point error in determination of $J_\ell(x)$ throws Newton's method into a loop for starting points too close to the origin. Thus, the functions built into SciPy are useless if they use $x=0$ as a starting point in Newton's algorithm for $\ell$ any larger than about 200. On our machine, the first failure of these built-in methods occurs at $\ell=231$, far too small to compute the zeta function with any impressive accuracy. Ultimately, we have efficiently computed roots for $\ell \lesssim 10,000$.\\

The problem would be alleviated if a better starting position were known than $x=0$. In fact, the first root of $J_\ell(x)$ is always at least $\ell$, but we can do much better than this. The roots of the Bessel functions are known to be interlaced. That is,

\begin{align}
j_{\ell-1,n} < j_{\ell,n} < j_{\ell-1,n+1}
\end{align}

\noindent so if one had already computed the roots to $J_{\ell-1}$, the work required to find those of $J_\ell$ is greatly reduced. Since all of the roots are necessary, this is the approach we took. This sequential computation was employed, and using the interlacing property described above, the Bisection method was found to be most efficient, converging usually within 20-30 iterations to the threshold accuracy of $10^{-10}$.\\

The number of roots $j_{\ell,n}$ less than some value $N$ scales as $\mathcal{O}(N^2)$, so multiprocessing was applied to cut down on computation time. The computation for $N=10^5$ was preformed on a 2015-era laptop with a 4-core processor, and completed in about 90 minutes, averaging $\sim$4 ms per root found. For $N=10^6$ our machine would need to run for an estimated 6 days, prohibitively long.\\

\noindent The result of the computation, shown as a density plot of the $j_{\ell,n}$ for bins of size 1 is shown below in Figure 1. The density is consistent asymptotically with the function:

\begin{align}
    \rho(j) \sim \frac{j}{2}-\frac{1}{4} + \text{error}
\end{align}

\noindent The error between this data and the asymptotic form $\rho(j)$ is randomly distributed around zero and seems to grow in absolute value no faster than $\sqrt{j}\times \text{constant}$. Within these relative bounds on the error, the roots are not distributed uniformly. 

\begin{center}
\begin{figure}[h!]
  \includegraphics[width=1\textwidth]{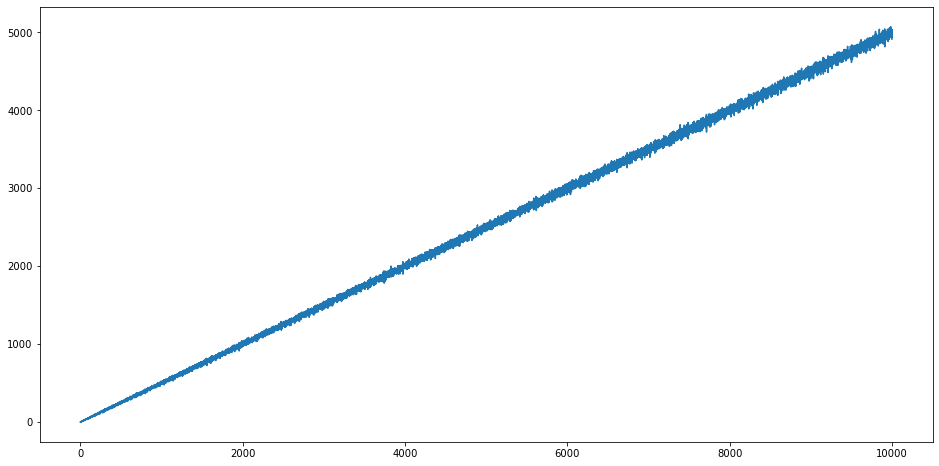}  \caption{Number of Bessel roots $j_{\ell,n}$ per unit interval, including only those less than $10^5$. Values for $\ell \neq 0$ have been counted twice, accounting for the 2-fold degeneracy of waves in 2+1 dimensions.}
\end{figure} 
\end{center}

Knowing the asymptotic form of this eigenvalue distribution is important. The summations involved in computing the zeta function converge slowly relative to those of the Riemann zeta function, so the function $\rho(j)$ allows us to approximate the remainder of the infinite sums without actually generating the roots themselves. Particular values of the zeta function have been found analytically\cite{1}. For instance:
\begin{align}
\zeta_\odot(4) = \sum_{\ell=-\infty}^\infty \sum_{n=1}^\infty \frac{1}{j^{4}_{\ell,n}} = \frac{2\pi^2-15}{96} \approx 0.049366758356028
\end{align}

\noindent Direct computation of this sum, cut off at $j_{\ell,n} \leq 10^5$, with accuracy of each root to 10 decimal places, yields

\begin{align}
\sum_{j \leq 10^5} \frac{1}{j^{4}_{\ell,n}} \approx 0.049366755856...
\end{align}

\noindent which is accurate for the first 8 decimal places. This can be improved significantly by simply adding to the sum an integration term:

\begin{align}
\sum_{j \leq 10^5} \frac{1}{j^{4}_{\ell,n}} +\int_{10^5}^\infty \frac{\rho(j)}{j^4}dj \approx 0.049366758356...
\end{align}

\noindent thereby increasing the accuracy to 12 decimal places. In the next section we characterize how these inaccuracies build when analytic continuation is preformed.\\

\subsection{Analytical Continuation}

For simplicity, we solved the discrete Cauchy-Riemann equations on a square lattice of spacing $\varepsilon$ representing the complex plane. This requires a connected domain of initial values for the zeta function, which were acquired by directly computing the summation described in section 2.1, with the integral over $\rho(j)$ to represent the zeros which were not explicitly derived. This procedure does not work well near poles, but since we know where the poles are and their residues, this can be avoided. Strictly speaking, we computed functions $u$ and $v$:

\begin{align}
    \varphi(s) = \zeta_\odot(s)-\frac{1}{2(s-2)}+\frac{1}{2(s-1)}+\frac{1}{128(s+1)} \equiv u+iv
\end{align}

\noindent (which is analytic for $\text{Re}(s) > -3$) then iteratively solved the discrete Cauchy-Riemann equations
\begin{align}
    u_{i,j} &= u_{i+2,j}-v_{i+1,j+1}+v_{i+1,j-1}\\
    v_{i,j} &= v_{i+2,j}+u_{i+1,j+1}-u_{i+1,j-1}
\end{align}

\noindent where the index $i$ represents steps in the positive real direction of the complex $s$ plane, and the index $j$ represents steps in the positive imaginary direction. These formula easily compute a region to the left of your initial known values, and can be manipulated to compute regions above, below, and to the right. Our procedure was as follows: directly compute the zeta function for a vertical strip of width 2$\varepsilon$ and height $h$ (symmetric about the real axis), then apply the Cauchy-Riemann equations to estimate the zeta function on a domain to the left and to the right of this strip. The final domain of computed values spans a $45^\circ$ rotated square in the complex plane, centered on $s=x_0+i0$, with corners at $x_0-h/2 + i0, x_0+h/2 +i0, x_0 +ih/2, x_0-ih/2$.\\

The purpose of computing for larger real values of $s$ was to establish bounds on the failure of this iterative process to accurately compute $\zeta_\odot$. To this end, $\varepsilon$, $h$, and $x_0$ had to be chosen wisely. Too many steps (too small $\varepsilon$) resulted in the blowup of floating point errors, and too few steps created errors which we believe represent the failure of coarse-graining of the Cauchy-Riemann equations. $x_0$ needed to be as close to $s=-1$ as possible without incurring the wrath of the simple pole at $s=2$ (the summation does not converge well at all for $s\lesssim 4+i0$), but not too large lest we run into the problem of step size.\\

Ultimately, we settled on the parameters $x_0 = 8$, $\varepsilon = 0.36$, $h=36$. $h$ was chosen to be so large so that we were guaranteed to see a blowup of the algorithm at some definite number of steps. We found this to be a good combination of the described constraints, and this is apparent in the data. In Figure 2 you will find the numerically generated plots of $\varphi(s)$ on the real axis.\\

\begin{center}
\begin{figure}[h!]
  \includegraphics[width=1\textwidth]{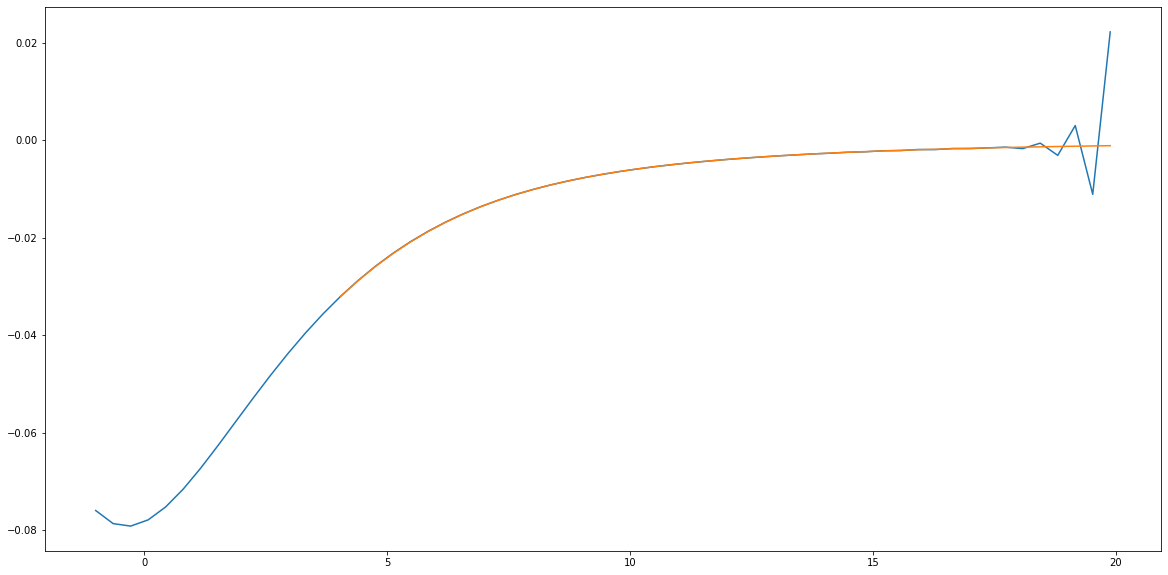}
  \includegraphics[width=1\textwidth]{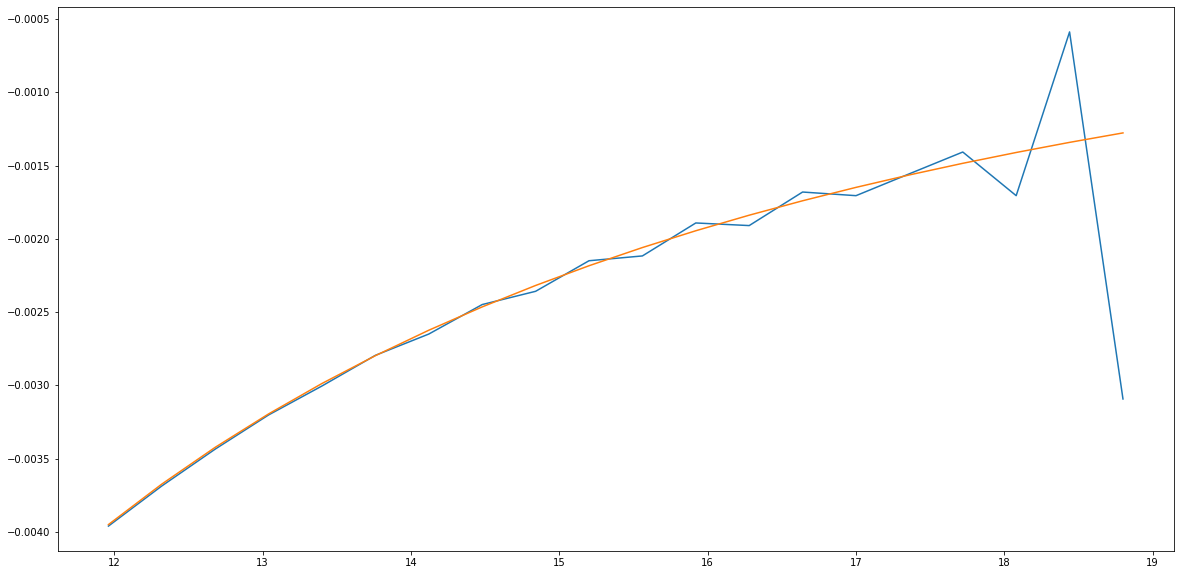}\caption{In blue, we show the numerically determined (analytically continued) values of $\varphi(s)$ along the real axis, and in orange, the exact determination using the summation described in section 2.1. A portion of the entire domain,  $x \in (-1,20)$ is shown (top) as well as a zoomed-in section $x\in(12,19)$ used to illustrate the growth of numerical error (bottom).}
\end{figure}     
\end{center}

In Figure 2, we see that the function $\varphi(s)$ is negative, but beginning to turn around near $s=-1$. This is consistent with the existence of a simple pole of positive residue at $s=-3$, which is known to be the case. As previously stated, we extracted a value of $\varphi(-1) \approx -0.07594$, and estimate a relative error of about 5\% for this value. This estimate is determined by inspecting the relative (or absolute) error at the location $s=17$, which is equal number of lattice steps to the right of our starting point $x_0$ as the point of interest, $s=-1$. By $s=19$, the error build up becomes large, and this is also seen to the left of the graph, close to the pole at $s=-3$. However, due to the pole at this location, we did not feel displaying the calculated values was of any interest, as the values are quite large. At the location $s=17$, the values of the blue and orange graphs are $-0.001706$ and $-0.001649$, respectively. The 5\% error bound on $\varphi(-1)$ squarely places the renormalized value of $\zeta_{\odot,\text{ren}}(-1) \approx 0.0074\pm0.0038$, which is positive.

\section{Implications for Physical Defects at Strong Coupling}

\subsection{Domain Walls in 2+1 dimensional Scalar Theories}

For simplicity, consider the 2+1 dimensional scalar $\varphi^4$ theory. We will call the mass of the quanta $m$ and the quartic coupling term (which has dimensions of mass) $\lambda$. When the potential possesses two degenerate vacuua, there are infinite linear defects (domain walls) in the model even at the classical level; these are the 2 dimensional ``kinks" and they have tension proportional to $T \sim m^3/\lambda$. That extended objects should have tensions inversely proportional to the underlying field theories coupling constants is a generic feature\cite{3}. The width of the domain walls are independent of $\lambda$ at classical order, and scale with the inverse mass $w \sim m^{-1}$. The quantum field theory is super-renormalizable, as the beta function for $\lambda$ vanishes. We can, therefore, consider the scenario where $\lambda$ is large compared with $m$, and compare the classical tension with first order quantum corrections. There is of course no guarantee that higher order quantum corrections can be discarded, but as a proof of concept the Casimir effect alone will serve well.\\

Consider also the possibility of a dilaton in the spectrum of our contrived model. The effective action of the dilaton may include couplings to the geometric invariants, but also to operators constructed of $\varphi$ and its derivatives. Since the energy and pressure is localized on the domain wall, the dilaton may be effectively considered a free particle with boundary conditions on the wall. We may now consider the implications if those boundary conditions are (at least approximately) Dirichlet.\\

For a compact, circular domain wall, the total energy of the system, including classical and first order quantum corrections, is therefore

\begin{align}
    E(a) = 2\pi aT + \frac{\zeta_{\odot,\text{ren}}(-1)}{a}
\end{align}

\noindent which has a global minima at 

\begin{align}
    a_* = \sqrt\frac{\zeta_{\odot,\text{ren}}(-1)}{2\pi T}
\end{align}

\noindent and minimal total energy

\begin{align}
    E(a_*) = \sqrt{4\pi \zeta_{\odot,\text{ren}}(-1) T}
\end{align}

It is important that we require the size of the wall to be large compared to its thickness, so that the Dirichlet boundary conditions may be applied appropriately for the IR modes of the dilaton. To this end, we must have

\begin{align}
    a_* m >> 1 \quad \text{or}\\
    \frac{\lambda}{m} >> \frac{2\pi}{\zeta_{\odot,\text{ren}}(-1)}
\end{align}

\noindent The right hand side of the second inequality is approximately 1000, so  $m$ must be quite small compared to $\lambda$ for this to work. Such a fine tuning can occur if we are near the critical point of the symmetry breaking phase transition of the field $\varphi$. $E(a_*)$ is also parametrically small compared to $m$ in this regime, indicating a stability against decay into the heavier $\varphi$ quanta. We therefore conclude the dilatons to be responsible for a Casimir force that stabilizes compact domain walls in this model, either very near to the critical point (after the phase transition), or for situations where $\lambda$ is abnormally large. We make no hypothesis of the circumstances which lead to this situation, although we do stress the result is indicative of a general phenomena of stabilization of classically forbidden solitons at strong coupling. Perhaps with a more detailed analysis of higher order corrections, the factor of 1000 will be rounded down significantly.\\

\subsection{Axion Membranes}

Axions are the Goldstone Bosons of a broken global $U(1)$ symmetry at very high energies. At low energies, the Axion acquires a mass, due to assumed gauge anomalies. The QCD Axion specifically becomes massive due to instanton condensation: correlators of $\text{Tr}(F\tilde{F})$ are non-vanishing\cite{4}. Regardless of the specifics of the Axion model, whether it is the exact QCD Axion or just Axion-like models, extended physical defects appear in the spectrum. The non-trivial winding of the Axion field around the vacuum manifold of the $U(1)$ produces objects called Cosmic/Axion/Global strings\cite{5}. Classically, the string tension is divergent, unless another string with opposite winding number screens the long distance effects at infinite volume. We have previously demonstrated that first order quantum corrections to the global string tension may serve to cancel this divergence at strong coupling\cite{6}. We therefore will ignore this effect.\\

When the Axion field anomalously acquires a mass, the degeneracy of the vacuum manifold is lifted and a membrane forms that connects different components of spatially separated strings\cite{4}. We will refer to the (finite) string and membrane tensions as $T$ and $\Sigma$, respectively. We will call the anomalous mass of the Axion $m_a$, and note that the width of the membrane is of order $m_a^{-1}$.\\

If we suppose a circular string is formed, and later its membrane, then we may calculate the total energy in the regime where membrane excitations are allowed but the boundary string is fixed in place. This is the case where the energy stored in the string is much greater than that of the membrane. Since the membrane oscillations have the same energy density as tension, the waves are massless and described by the spectrum of the ordinary Bessel functions $J_\ell$. Therefore the total energy is

\begin{align}
    E(a) = 2\pi a T + \pi a^2 \Sigma +\frac{\zeta_{\odot,\text{ren}}(-1)}{a}
\end{align}

We however assume that the second term is much smaller than the first and third components. In other words

\begin{align}
    a << \frac{2\pi T}{\Sigma}
\end{align}

\noindent The solution is, as before, what we call $a_*$, and it must be large compared to $m_a^{-1}$. Denote the scale of the $U(1)$ symmetry breaking by $M_\text{SSB}$ and the coupling constant of this field by $\lambda$. The tension $\Sigma$ scales with $m_a$ like $\Sigma \sim m_a f^2_a$, where $f_a$ is the scale of physics that generates the anomaly (for the QCD Axion, $M_\text{SSB} = f_a$, the Axion decay constant)\cite{5}. In principle there is no reason these cannot be different scales for generic Axion models. The heirarchy of scales is therefore:

\begin{align}
    1 << \frac{m_a}{M_\text{SSB}}\sqrt{\frac{\lambda \zeta_{\odot,\text{ren}}(-1)}{2\pi}} << 2\left(\frac{M_\text{SSB}}{f_a}\right)^2
\end{align}

\noindent The first inequality is the condition that the thin-membrane approximation is valid at the stable radius, and the second inequality the condition that membrane oscillations are the primary source of the Casimir effect, or that oscillations of the string boundary are kinetically inaccessible. These are the approximations we must take to apply the results of our Zeta function calculation with confidence.\\

Our thin wall approximation is easily met if $\lambda$ is parametricaly large compared to the mass ratio of the heavy $U(1)$ quanta to the anomalous Axion mass. $\lambda$ can be small if $m_a >> M_\text{SSB}$, but this is likely prohibited phenomenologically, at least in the case that the Axion is weakly coupled with the anomalous physics. For the QCD Axion, these inequalities clearly cannot be met, and an analysis must be done of the full vibration spectrum of the string+membrane system (much larger radii). This involves determination of the zeta function for the roots of Bessel functions, $j^\prime_{\ell,n}$. We attempted this calculation but progress was slow since the residues and locations of the poles are unknown to this author. We expect the result to be repulsive, as the exchange of Dirichlet to Neumann boundary conditions does not significantly effect the spectral density.\\

\section{Conclusion}

We have demonstrated that the Casimir forces on 2 dimensional circular or disk-like objects are repulsive, when renormalized in the MS/pole subtraction scheme. Such a subtraction is warranted as the problem is in the UV, not the IR. Consequently, there are quantum modifications to the total energy of solitons in this class of geometries. Of particular interest is the membrane formed in the vicinity of Axion strings, after the Axion receives a small anomalous mass. Generically, these repulsive forces may produce stable, non-contracting solitons if there is an underlying coupling constant which is large. The exact results in sections 3.1 and 3.2 are not to be taken literally, but rather as evidence that quantum mechanical effects may in fact stabilize these objects when coupling constants are large. We conjecture that a systematic calculation of higher order effects will reveal that the coupling need only be order unity for the conclusions of this section to hold in some regime.\\

\end{document}